\pdfoutput=1
\documentclass[aps,prl,twocolumn,showpacs,superscriptaddress,a4paper,
    floatfix]{revtex4}
\usepackage[latin1]{inputenc}
\usepackage{graphicx}
\usepackage{amsmath,amssymb}
\usepackage{hyperref}
\usepackage{datetime}

\begin{document}

\title{Superfluidity with disorder in a quantum gas thin film}

\author{Sebastian Krinner}
\affiliation{Department of Physics, ETH Zurich, 8093 Zurich,
Switzerland}
\author{David Stadler}
\affiliation{Department of Physics, ETH Zurich, 8093 Zurich,
Switzerland}
\author{Jakob Meineke}
\affiliation{Department of Physics, ETH Zurich, 8093 Zurich,
Switzerland}
\author{Jean-Philippe Brantut}
\email{brantutj@phys.ethz.ch} 
\affiliation{Department of Physics, ETH Zurich, 8093 Zurich,
Switzerland}
\author{Tilman Esslinger}
\affiliation{Department of Physics, ETH Zurich, 8093 Zurich,
Switzerland}

\date{\pdfdate}

\begin{abstract}

We investigate the properties of a strongly interacting, superfluid gas of $^6$Li$_2$ Feshbach molecules forming a thin film confined in a quasi two-dimensional channel with a tunable random potential, creating a microscopic disorder. We measure the atomic current and extract the resistance of the film in a two-terminal configuration, and identify a superfluid state at low disorder strength, which evolves into a normal, poorly conducting state for strong disorder. The transition takes place when the chemical potential reaches the percolation threshold of the disorder. The evolution of the conduction properties contrasts with the smooth behavior of the density and compressibility across the transition, measured in-situ at equilibrium. These features suggest the emergence of a glass-like phase at strong disorder. 

\end{abstract}


\maketitle

Disorder has profound effects on the properties of materials. It can turn metals \cite{RevModPhys.73.251}, or even superconductors and superfluids into insulators \cite{goldman:39,springerlink:10.1007/BF00114905,PhysRevB.55.12620}. The evolution from a superfluid state to an insulator depends on the details of the underlying material structure, which is complex and often only partially known. Several models have been devised, accounting for various aspects of the transition \cite{1063-7869-53-1-R01}. Dirty boson models, where interacting bosons are placed in a random potential, have attracted particular attention. These models predict new phases such as the Bose-glass, expected as a result of disorder and interactions \cite{PhysRevB.40.546,PhysRevLett.103.140402,Aleiner:2010kx}. 

Over the last years, ultracold gases in random potentials have emerged as an important tool to investigate the interplay between disorder and superfluidity. They offer comparatively simple systems controlled by a few microscopic parameters that are known and controlled \cite{1751-8121-45-14-143001}. They have been studied using observables such as phase coherence, response to external forces and lattice modulations \cite{PhysRevLett.98.130404,2009PhRvL.102e5301W,Pasienski:2010vn,PhysRevLett.107.145306,PhysRevA.85.033602,1367-2630-14-7-073024}, but transport coefficients have not been measured, preventing a direct comparison with condensed matter systems.

In this letter, we present measurements of the DC resistance of a disordered, strongly interacting gas of $^6$Li$_2$ molecules. We use the recently demonstrated two-terminal setup \cite{Brantut31082012,Stadler:2012vn} to measure the resistance of a quasi two-dimensional thin film of molecules as a function of disorder strength, in a manner directly analogous to solid-state experiments. By comparing the resistance of the strongly interacting gas of molecules to that of a weakly interacting Fermi gas in the same potential landscape, we observe a breakdown of superfluid flow when the percolation threshold of the disorder reaches the chemical potential of the gas. In addition, we measure the equilibrium density and compressibility of the disordered gas, which do not display large variations up to very strong disorder, a feature expected for a glass transition.

\begin{figure}[h!tb]
    \includegraphics[width=0.4\textwidth]{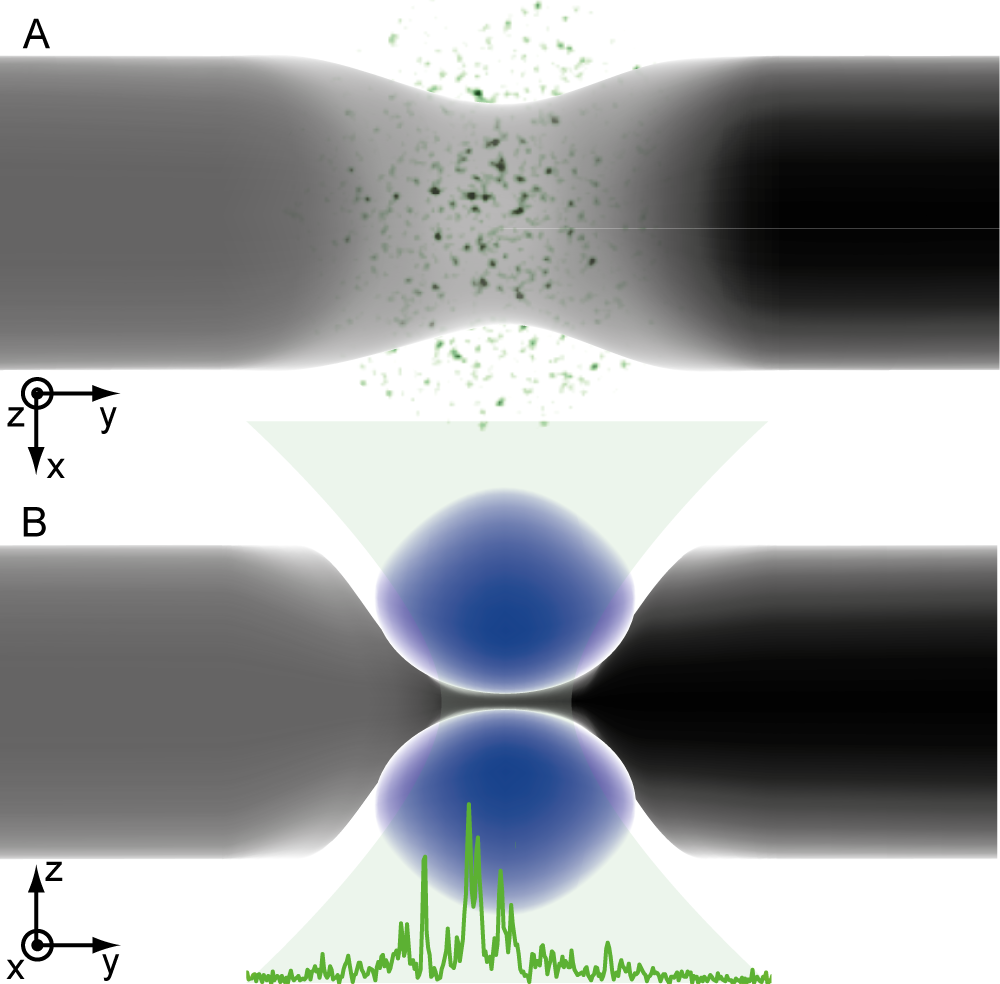}
    \caption{Sketch of the experimental configuration. A: top view. B: side view. Two $^6$Li$_2$ Feshbach molecules reservoirs (grey) are connected by a narrow, quasi-2D channel, created by a laser beam (in blue) at 532\,nm with a $1/e^2$-waist of $30$\,$\mu$m along $y$, having a nodal line at the center. When a number imbalance between the reservoirs is present, a molecule current sets in through the channel. A speckle pattern (green) of variable intensity is projected onto the channel along the $z$-axis through a microscope objective (not shown). }
    \label{fig:fig1}
\end{figure}

The experimental setup is based on our previous work \cite{Brantut31082012}, and is sketched in figure \ref{fig:fig1}. We prepare a cloud of molecules formed during evaporative cooling of a $^6$Li gas in a mixture of the two lowest hyperfine states at a magnetic field of 751\,G. At this field, the s-wave scattering length is $3545\,a_0$, where $a_0$ is Bohr's radius, and the binding energy of the molecules is $E_b = 2.3\,\mu$K \cite{PhysRevLett.94.103201}. The elongated cloud of $1.03(5)\,10^5$ molecules \cite{errorbars} is separated at its center into two reservoirs by a repulsive potential created with a laser beam having a nodal line at its center. 
This creates a quasi-two-dimensional channel which connects the two reservoirs. The trap frequency along the tightly confining $z$-axis at the center of the channel is $\omega_z=2 \pi \cdot  6.1$\,kHz, thus $\hbar \omega_z \ll E_b$.

Without the channel, the chemical potential of the molecular gas is $\mu_0=550(70)$\,nK, deduced from the particle number and trap frequencies \cite{kfa}.
The channel covers a region of $30$\,$\mu$m along $y$, small compared to the size of the cloud ($\simeq 300\,\mu$m). Therefore we assume that the overall chemical potential remains constant in the presence of the channel. Accordingly, the local chemical potential in the channel is $\mu = \mu_0 - \frac{1}{2} \hbar \omega_z \simeq 400(70)$\,nK. With $\mu \simeq 1.4\,\hbar \omega_z$, the gas predominantly populates the lowest vibrational state. The inter-molecule scattering length is $a_m = 2100\,a_0$ \cite{PhysRevLett.93.090404}, which yields an interaction parameter for the 2D scattering problem $\sqrt{8 \pi} a_m / l_{z} \simeq 1.1$, with $l_{z}$ being the harmonic oscillator length in the tightly confined direction \cite{PhysRevA.64.012706}. The molecular gas is thus in the strongly interacting regime, where the scattering amplitude depends on momentum \cite{PhysRevLett.84.2551}.

The disordered potential is realized by projecting a speckle pattern onto the channel through a high-numerical aperture microscope. 
A gaussian fit to the envelope of the pattern yields a $1/e^2$ radius of $36\,\mu$m. 
The average disorder strength $\bar{V}$ at the center of the pattern is calculated from the beam envelope and the laser power. 
The correlation length $\sigma$ of the disorder, defined by the $1/e$ radius of the autocorrelation function is $290(90)$\,nm.
Associated with this length scale, we introduce a correlation energy $E_{\sigma}=\frac{\hbar^2}{m\sigma^2}=0.48$\,$\mu$K, where $m$ is the mass of a molecule. This scale separates the quantum regime $\bar{V} \ll E_{\sigma}$ from a classical regime $\bar{V} \gg E_{\sigma}$ \cite{springerlink:10.1134/S1063782608080058}. 

It is instructive to compare these scales to that of a Bose-Einstein condensate (BEC) of molecules. The chemical potential in the channel is independent of the disorder, and fixed by the unperturbed particle reservoirs.
The size of a molecule, given by the interatomic scattering length, is 190\,nm, comparable but smaller than the correlation length of the disorder, so we expect the molecules to remain bound for increasing disorder. 
The healing length of the molecular BEC, associated with the chemical potential, is typically $\xi \simeq 230$\,nm, of the order of the correlation length of the disorder. Thus, we expect the disorder to affect the many-body physics at the microscopic level \cite{PhysRevA.85.033602}.

\begin{figure}[h!tb]
    \includegraphics[width=0.45\textwidth]{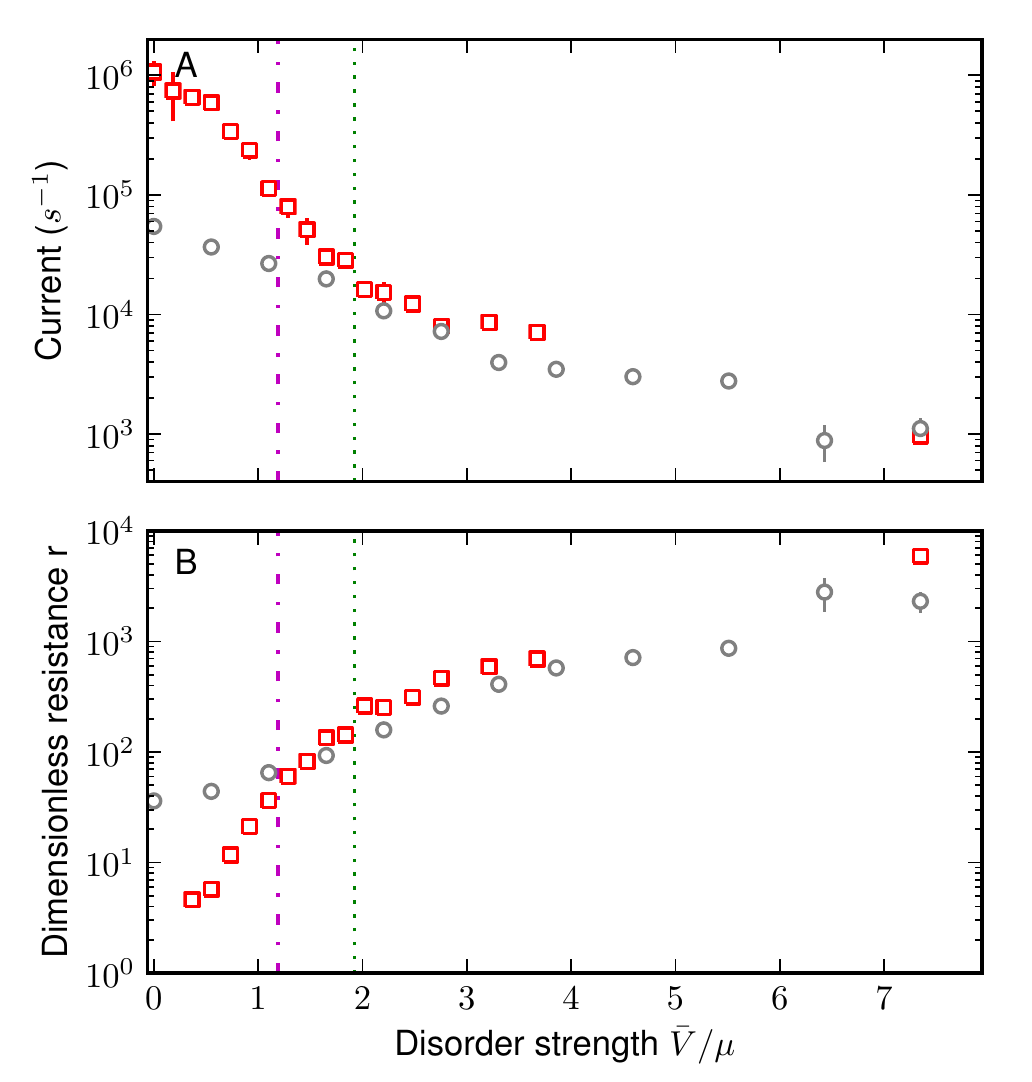}
    \caption{Conduction properties of the disordered gas of molecules (red squares), and of a weakly interacting Fermi gas (grey dots) as a function of disorder strength normalized by $\mu=400$\,nK, the chemical potential of the molecular gas in the channel without disorder. A: Initial current, for a fixed driving (see text). B: Dimensionless resistance. The dotted green line represents the classical percolation threshold of the speckle pattern. The dash-dotted line indicates the correlation energy $E_{\sigma}$.  Error bars are statistical.}
    \label{fig:currentsResistance}
\end{figure}

We investigate the conduction properties of the disordered gas by inducing a chemical potential difference between the two reservoirs \cite{Brantut31082012,Stadler:2012vn}. During evaporative cooling, a fixed relative atom number imbalance is created between the left and right reservoir. We observe the evolution of the atom number imbalance while a quasi-stationary current flows through the channel. We measure the slope of the initial evolution, yielding the maximum current driven by the chemical potential difference. Figure \ref{fig:currentsResistance}A shows the initial current as a function of $\bar{V}$. For weak disorder, we observe a large current, limited by conservation of energy. We interpret this as a manifestation of superfluidity, as expected for a BEC of molecules \cite{2005Natur.435.1047Z}. With increasing disorder, the current quickly decreases, and the time evolution of the number imbalance is well described by an exponential. In this regime, we fit a decay time $\tau_{\mathrm{BEC}} = RC$, where $R$ is the resistance of the channel and $C = \partial N/\partial \mu$ is the $\bar{V}$-independent compressibility of the reservoirs. Figure \ref{fig:currentsResistance}B shows $r$ = $RC\omega_y$, where $\omega_y$ is the underlying trap frequency of the reservoirs along the transport direction \cite{Stadler:2012vn}. For weak disorder, we observe a fast increase of $r$ with disorder strength, up to $\bar{V}/\mu \sim 2$, and a correspondingly fast decrease of current. For strong disorder, the increase of resistance shows a weaker but again exponential dependency on the disorder strength.
The highest resistance plotted represents a slow down of the transport by about three orders of magnitude compared to the case without disorder. It represents the lowest current and highest resistance we can measure.

To disentangle the effects of superfluidity from the single particle effects caused by disorder, we repeat the experiment in the same trap and disorder configuration with a weakly interacting Fermi gas (WIF) of $9.3(4)\,10^4$ atoms per spin state, prepared at a magnetic field of $475$\,G, where the scattering length is $-100$\,a$_0$. The current and resistance are shown as grey dots on figure \ref{fig:currentsResistance}. They show a single exponential dependency and a finite resistance at low disorder strength, corresponding to the contact resistance of the channel \cite{Brantut31082012}. This behavior confirms that the disorder strongly affects transport already at the single particle level \cite{2007NJPh....9..161K,PhysRevLett.104.220602,2011NJPh...13i5015P}.

We then extract a parameter $B = R_{\mathrm{BEC}}/R_{\mathrm{WIF}}$, which measures the effect of pairing and superfluidity compared to the ideal Fermi gas case, keeping the geometry and potential landscape unchanged. It accounts, at least qualitatively, for the variations of resistance in a non superfluid case, without relying on a model for the complicated single particle dynamics. This procedure resembles the use of the normal state resistance as a reference in the physics of superconductors. We evaluate $B=K \tau_{\mathrm{BEC}}/\tau_{\mathrm{WIF}}$ where $K$ is the ratio of the compressibility of the reservoirs \cite{sfreservoirs}, which does not depend on disorder strength. Modeling the reservoirs as harmonically trapped zero temperature Fermi gases and weakly interacting Bose-Einstein condensate, we estimate $K \simeq 0.57$.

\begin{figure}[h!tb]
    \includegraphics[width=0.45\textwidth]{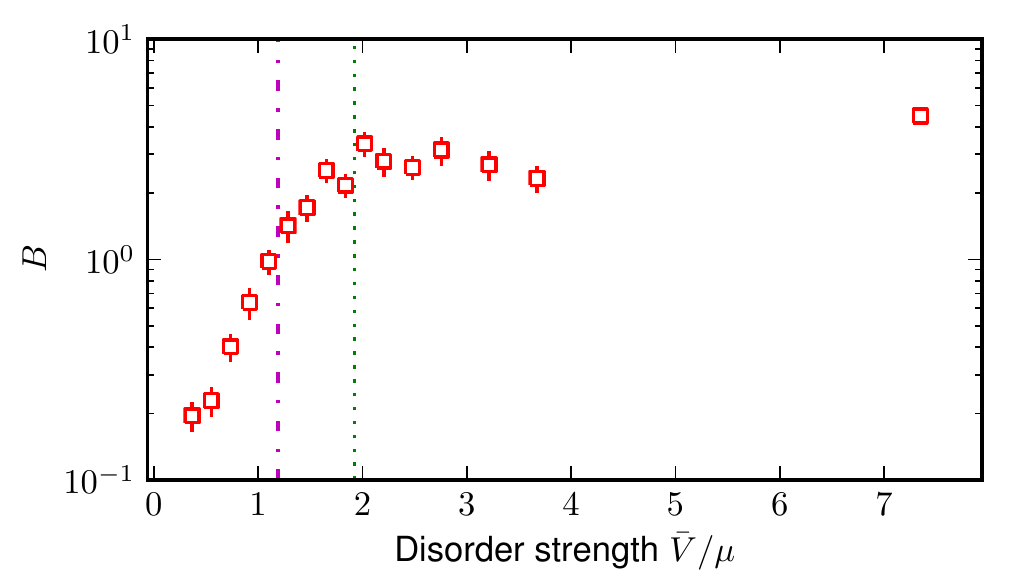}
    \caption{$B$ parameter (see text) as a function of disorder strength.
    The green dotted line represents the percolation threshold of the speckle potential, and coincides with the change of behavior from superfluid to single-particle transport. The dash-dotted line shows the correlation energy $E_{\sigma}$. Error bars represent statistical errors.}
    \label{fig:beta}
\end{figure}

Figure \ref{fig:beta} shows the evolution of $B$ with disorder strength, obtained by the ratio of the decay times for the molecules and the linearly interpolated data of the WIF. For strong disorder, it remains constant. The exact value should depend on the chemical potentials of the reservoirs for the two cases, but it should reflect the zero temperature behavior of conductance \cite{tempeffects}. This indicates that transport in the strongly disordered regime is dominated by single atom or molecule physics \cite{2007NJPh....9..161K}. In contrast, for weak disorder, $B$ varies very quickly, incompatible with simple diffusion \cite{2007NJPh....9..161K}. $B$ displays a sharp change of behavior for intermediate disorder strength, at the point where the chemical potential of the reservoirs reaches the percolation threshold of the disorder  $\bar{V}_{t} = 1.92 \mu$ \cite{PhysRevB.20.3653,PhysRevB.26.1352} (green dotted line). This corresponds to the intuitive picture of a transition from isolated superfluid pockets to a connected superfluid allowing for superflow. It agrees with the concept of a Bose-glass composed of an incoherent ensemble of localized condensates \cite{2007PhRvL..98q0403L,1367-2630-12-7-073003, springerlink:10.1134/S1063782608080058,PhysRevB.80.104515,Bourdel:2012fk}. However, contrary to the superfluid to Bose-glass quantum phase transition, the behavior in the strongly disordered case resembles that of a metal with finite conductance. Since the Bose-glass is smoothly connected at finite temperature to the normal Bose gas, we propose that the finite resistance is a finite temperature effect. It is remarkable that the classical picture of percolation gives a quantitative description of our observations even though interactions are strong and tunneling through individual disorder grains is significant, as $\bar{V}_{t}\sim 1.6 E_{\sigma}$ \cite{springerlink:10.1134/S1063782608080058}.

We now investigate how the density and compressibility in the channel depend on the disorder strength. We prepare a cloud with equal population in the two reservoirs. The disorder is switched on and an absorption picture is taken in-situ after $150$\,ms, allowing the channel to thermalize with the reservoirs \cite{timescale}. We typically average 20 of those pictures to reduce noise. We first consider the density observed at the center of the channel, averaged over a region of $18$\,$\mu$m ($7$\,$\mu$m) along the $y$ ($x$) direction. Over this region, the trapping frequency along $z$ varies by less than $10$\,\%, and the local chemical potential of the cloud by about $30$\,\%. The area of this region is much larger than the correlation length of the disorder, therefore we expect the spatial average to reflect the average over disorder realizations. 

\begin{figure}[h!tb]
    \includegraphics[width=0.45\textwidth]{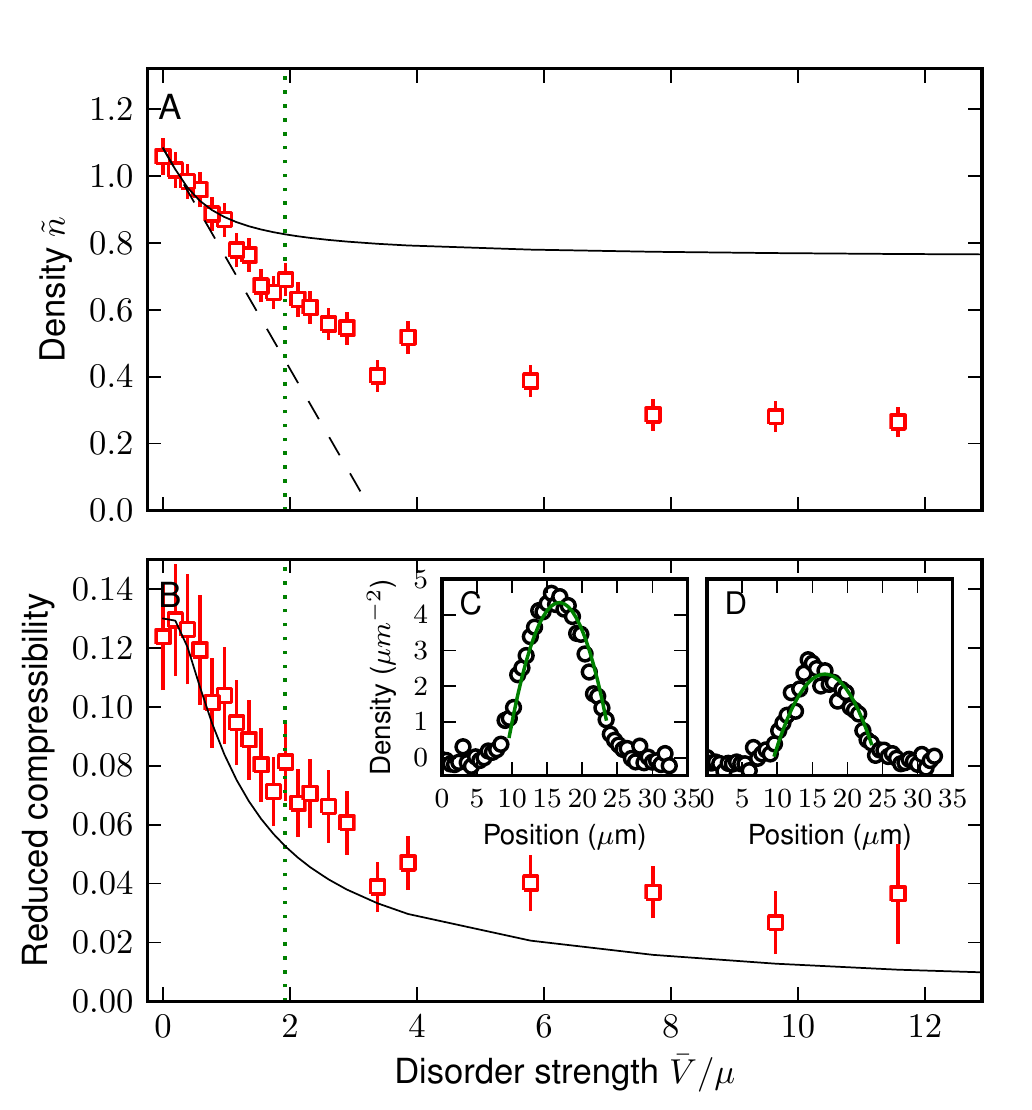}
    \caption{Equilibrium properties of the disordered gas as a function of disorder strength. The green dotted line indicates the percolation threshold. The solid black line indicates the prediction of the Thomas-Fermi approximation. A: average density at the center of the channel. The dashed line indicates the expected decrease of density from the compressibility measurement without speckle (see text). B: Reduced compressibility (see text) estimated from the shape of the cloud. Insets: density profiles in the channel (open circles) and quadratic fit to the central part (continuous line) used to extract the compressibility, for $\bar{V}/\mu=0$ and $1.6$ (C and D).}
    \label{fig:fig3}
\end{figure}

Figure \ref{fig:fig3}A presents the number of atoms per correlation area of the disorder $\tilde{n}=n  \pi \sigma^2$ as a function of disorder strength. The density smoothly decreases with increasing disorder up to the highest disorder strength. This is expected since the disorder is repulsive and reduces the available phase space. 
The density changes by a factor of 5, which is small compared to the three orders of magnitude change in the resistance. Therefore, the fast variation in resistance can not be solely attributed to a change in density.

We now make use of the in-situ density measurements to evaluate the compressibility. Because the trapping potential along $x$ varies much slower than the correlation length of the disorder, we use the local density approximation to obtain the variation of the disorder-averaged density with chemical potential, i.e. the disorder-averaged compressibility $d\bar{n}/d\mu$, where $\bar{n}$ is the disorder-averaged density. At the center of the cloud, the curvature of the density distribution reads \cite{2009Natur.460..995G}
$
\frac{d^2 \bar{n}}{dx^2} = - m \omega_x^2 \frac{d \bar{n}}{d\mu},
$
where we include in $\omega_x$ the small effect of the envelope of the speckle pattern. We use a quadratic fit to the central part of the cloud to extract the curvature of the density profile, shown on figure \ref{fig:fig3}C and D for two different disorder strength.

Figure \ref{fig:fig3}B presents the reduced compressibility $\frac{\hbar^2}{m} \frac{d\bar{n}}{d\mu}$. 
For very weak disorder, the compressibility remains constant, equal to that of the system without disorder. For stronger disorder, it decreases continuously with disorder strength. The overall variation remains within a factor of $\sim 5$. The compressibility remains finite up to very strong disorder where transport is frozen, suggesting a glassy phase. 

The zeroth order effect of a weak repulsive disorder is a shift of the average chemical potential equal to the mean disorder strength. We can use this and the measured compressibility to estimate the decay of the density due to the disorder (dashed line in figure \ref{fig:fig3}A). The agreement gives us confidence in the compressibility measurement.

We can further compare these observations to the predictions of a zero-temperature Thomas-Fermi approximation for a random potential \cite{Bourdel:2012fk}, shown as a black line in figure \ref{fig:fig3}A (B). The free parameters are the chemical potential and compressibility in absence of disorder. We find good agreement at low disorder, where the disorder mainly gives an overall shift of $\bar{V}$ of the chemical potential. For higher disorder, the observed density lies below the predictions. This is expected as $\tilde{n}<1$ and interactions are strong, so that the fragments of BEC are strongly depleted. Surprisingly, the shape of the compressibility data is well reproduced. 

Our measurements have shown the difference between equilibrium and transport properties of a superfluid bosonic thin film with disorder. 
By tuning the interaction using the Feshbach resonance, our setup could be used to investigate another, Fermionic scenario, where pairing results from a many-body process and can be strongly affected by the disorder \cite{PhysRevLett.99.250402,1367-2630-13-5-055012}. 

We acknowledge fruitful discussions with J. Blatter, T. Bourdel, T. Donner, V. Josse, P. Lugan, C. Mueller, L. Pollet, T. Roscilde, D. Shahar, V. Shenoy, A. Zheludev and W. Zwerger. We acknowledge financing from NCCR MaNEP and QSIT, ERC project SQMS, FP7 project NAME-QUAM and ETHZ. JPB is supported by the EU through a Marie Curie Fellowship.


\begin{thebibliography}{10}
\providecommand{\bibnamefont}[1]{#1}
\providecommand{\bibfnamefont}[1]{#1}
\providecommand{\bibinfo}[2]{#2}
\providecommand{\biburlhref}[2]{\href{#1}{#2}}

\bibitem{RevModPhys.73.251}

\bibinfo{author}{\bibfnamefont{E.}~\bibnamefont{Abrahams}},
  \bibinfo{author}{\bibfnamefont{S.~V.} \bibnamefont{Kravchenko}},
  \bibnamefont{and} \bibinfo{author}{\bibfnamefont{M.~P.}
  \bibnamefont{Sarachik}}, \bibinfo{journal}{Rev. Mod. Phys.}
  \textbf{\bibinfo{volume}{73}}, \bibinfo{pages}{251} (\bibinfo{year}{2001}).

\bibitem{goldman:39}

\bibinfo{author}{\bibfnamefont{A.~M.} \bibnamefont{Goldman}} \bibnamefont{and}
  \bibinfo{author}{\bibfnamefont{N.}~\bibnamefont{Markovic}},
  \bibinfo{journal}{Physics Today} \textbf{\bibinfo{volume}{51}},
  \bibinfo{pages}{39} (\bibinfo{year}{1998}).

\bibitem{springerlink:10.1007/BF00114905}

\bibinfo{author}{\bibfnamefont{J.~D.} \bibnamefont{Reppy}},
  \bibinfo{journal}{Journal of Low Temperature Physics}
  \textbf{\bibinfo{volume}{87}}, \bibinfo{pages}{205} (\bibinfo{year}{1992}).

\bibitem{PhysRevB.55.12620}

\bibinfo{author}{\bibfnamefont{P.~A.} \bibnamefont{Crowell}},
  \bibinfo{author}{\bibfnamefont{F.~W.} \bibnamefont{Van~Keuls}},
  \bibnamefont{and} \bibinfo{author}{\bibfnamefont{J.~D.} \bibnamefont{Reppy}},
  \bibinfo{journal}{Phys. Rev. B} \textbf{\bibinfo{volume}{55}},
  \bibinfo{pages}{12620} (\bibinfo{year}{1997}).

\bibitem{1063-7869-53-1-R01}

\bibinfo{author}{\bibfnamefont{V.~F.} \bibnamefont{Gantmakher}}
  \bibnamefont{and} \bibinfo{author}{\bibfnamefont{V.~T.}
  \bibnamefont{Dolgopolov}}, \bibinfo{journal}{Physics-Uspekhi}
  \textbf{\bibinfo{volume}{53}}, \bibinfo{pages}{1} (\bibinfo{year}{2010}).

\bibitem{PhysRevB.40.546}

\bibinfo{author}{\bibfnamefont{M.~P.~A.} \bibnamefont{Fisher}}, \emph{et~al.},
  \bibinfo{journal}{Phys. Rev. B} \textbf{\bibinfo{volume}{40}},
  \bibinfo{pages}{546} (\bibinfo{year}{1989}).

\bibitem{PhysRevLett.103.140402}

\bibinfo{author}{\bibfnamefont{L.}~\bibnamefont{Pollet}}, \emph{et~al.},
  \bibinfo{journal}{Phys. Rev. Lett.} \textbf{\bibinfo{volume}{103}},
  \bibinfo{pages}{140402} (\bibinfo{year}{2009}).

\bibitem{Aleiner:2010kx}

\bibinfo{author}{\bibfnamefont{I.~L.} \bibnamefont{Aleiner}},
  \bibinfo{author}{\bibfnamefont{B.~L.} \bibnamefont{Altshuler}},
  \bibnamefont{and} \bibinfo{author}{\bibfnamefont{G.~V.}
  \bibnamefont{Shlyapnikov}}, \bibinfo{journal}{Nat Phys}
  \textbf{\bibinfo{volume}{6}}, \bibinfo{pages}{900} (\bibinfo{year}{2010}).

\bibitem{1751-8121-45-14-143001}

\bibinfo{author}{\bibfnamefont{B.}~\bibnamefont{Shapiro}},
  \bibinfo{journal}{Journal of Physics A: Mathematical and Theoretical}
  \textbf{\bibinfo{volume}{45}}, \bibinfo{pages}{143001}
  (\bibinfo{year}{2012}).

\bibitem{PhysRevLett.98.130404}

\bibinfo{author}{\bibfnamefont{L.}~\bibnamefont{Fallani}}, \emph{et~al.},
  \bibinfo{journal}{Phys. Rev. Lett.} \textbf{\bibinfo{volume}{98}},
  \bibinfo{pages}{130404} (\bibinfo{year}{2007}).

\bibitem{2009PhRvL.102e5301W}

\bibinfo{author}{\bibfnamefont{M.}~\bibnamefont{{White}}}, \emph{et~al.},
  \bibinfo{journal}{Physical Review Letters} \textbf{\bibinfo{volume}{102}},
  \bibinfo{pages}{055301} (\bibinfo{year}{2009}).

\bibitem{Pasienski:2010vn}

\bibinfo{author}{\bibfnamefont{M.}~\bibnamefont{Pasienski}}, \emph{et~al.},
  \bibinfo{journal}{Nat Phys} \textbf{\bibinfo{volume}{6}},
  \bibinfo{pages}{677} (\bibinfo{year}{2010}).

\bibitem{PhysRevLett.107.145306}

\bibinfo{author}{\bibfnamefont{B.}~\bibnamefont{Gadway}}, \emph{et~al.},
  \bibinfo{journal}{Phys. Rev. Lett.} \textbf{\bibinfo{volume}{107}},
  \bibinfo{pages}{145306} (\bibinfo{year}{2011}).

\bibitem{PhysRevA.85.033602}

\bibinfo{author}{\bibfnamefont{B.}~\bibnamefont{Allard}}, \emph{et~al.},
  \bibinfo{journal}{Phys. Rev. A} \textbf{\bibinfo{volume}{85}},
  \bibinfo{pages}{033602} (\bibinfo{year}{2012}).

\bibitem{1367-2630-14-7-073024}

\bibinfo{author}{\bibfnamefont{M.~C.} \bibnamefont{Beeler}}, \emph{et~al.},
  \bibinfo{journal}{New Journal of Physics} \textbf{\bibinfo{volume}{14}},
  \bibinfo{pages}{073024} (\bibinfo{year}{2012}).

\bibitem{Brantut31082012}

\bibinfo{author}{\bibfnamefont{J.-P.} \bibnamefont{Brantut}}, \emph{et~al.},
  \bibinfo{journal}{Science} \textbf{\bibinfo{volume}{337}},
  \bibinfo{pages}{1069} (\bibinfo{year}{2012}).

\bibitem{Stadler:2012vn}
\biburlhref{http://arxiv.org/abs/1210.1426}{%
\bibinfo{author}{\bibfnamefont{D.}~\bibnamefont{Stadler}}, \emph{et~al.},
  \bibinfo{journal}{arXiv e-print 1210.1426}  (\bibinfo{year}{2012}).}

\bibitem{PhysRevLett.94.103201}

\bibinfo{author}{\bibfnamefont{M.}~\bibnamefont{Bartenstein}}, \emph{et~al.},
  \bibinfo{journal}{Phys. Rev. Lett.} \textbf{\bibinfo{volume}{94}},
  \bibinfo{pages}{103201} (\bibinfo{year}{2005}).

\bibitem{errorbars}

\bibinfo{note}{Unless otherwise stated, error bars represent the combined
  statistical and estimated systematic errors.}

\bibitem{kfa}

\bibinfo{note}{In the reservoirs the interaction parameter is $k_F a \sim 1$,
  deep in the BEC regime. All the numbers in the text are stated for the
  molecules.}

\bibitem{PhysRevLett.93.090404}

\bibinfo{author}{\bibfnamefont{D.~S.} \bibnamefont{Petrov}},
  \bibinfo{author}{\bibfnamefont{C.}~\bibnamefont{Salomon}}, \bibnamefont{and}
  \bibinfo{author}{\bibfnamefont{G.~V.} \bibnamefont{Shlyapnikov}},
  \bibinfo{journal}{Phys. Rev. Lett.} \textbf{\bibinfo{volume}{93}},
  \bibinfo{pages}{090404} (\bibinfo{year}{2004}).

\bibitem{PhysRevA.64.012706}

\bibinfo{author}{\bibfnamefont{D.~S.} \bibnamefont{Petrov}} \bibnamefont{and}
  \bibinfo{author}{\bibfnamefont{G.~V.} \bibnamefont{Shlyapnikov}},
  \bibinfo{journal}{Phys. Rev. A} \textbf{\bibinfo{volume}{64}},
  \bibinfo{pages}{012706} (\bibinfo{year}{2001}).

\bibitem{PhysRevLett.84.2551}

\bibinfo{author}{\bibfnamefont{D.~S.} \bibnamefont{Petrov}},
  \bibinfo{author}{\bibfnamefont{M.}~\bibnamefont{Holzmann}}, \bibnamefont{and}
  \bibinfo{author}{\bibfnamefont{G.~V.} \bibnamefont{Shlyapnikov}},
  \bibinfo{journal}{Phys. Rev. Lett.} \textbf{\bibinfo{volume}{84}},
  \bibinfo{pages}{2551} (\bibinfo{year}{2000}).

\bibitem{springerlink:10.1134/S1063782608080058}

\bibinfo{author}{\bibfnamefont{B.}~\bibnamefont{Shklovskii}},
  \bibinfo{journal}{Semiconductors} \textbf{\bibinfo{volume}{42}},
  \bibinfo{pages}{909} (\bibinfo{year}{2008}).

\bibitem{2005Natur.435.1047Z}

\bibinfo{author}{\bibfnamefont{M.~W.} \bibnamefont{{Zwierlein}}},
  \emph{et~al.}, \bibinfo{journal}{Nature} \textbf{\bibinfo{volume}{435}},
  \bibinfo{pages}{1047} (\bibinfo{year}{2005}).

\bibitem{2007NJPh....9..161K}

\bibinfo{author}{\bibfnamefont{R.~C.} \bibnamefont{{Kuhn}}}, \emph{et~al.},
  \bibinfo{journal}{New Journal of Physics} \textbf{\bibinfo{volume}{9}},
  \bibinfo{pages}{161} (\bibinfo{year}{2007}).

\bibitem{PhysRevLett.104.220602}

\bibinfo{author}{\bibfnamefont{M.}~\bibnamefont{Robert-de Saint-Vincent}},
  \emph{et~al.}, \bibinfo{journal}{Phys. Rev. Lett.}
  \textbf{\bibinfo{volume}{104}}, \bibinfo{pages}{220602}
  (\bibinfo{year}{2010}).

\bibitem{2011NJPh...13i5015P}

\bibinfo{author}{\bibfnamefont{L.}~\bibnamefont{{Pezz{\'e}}}}, \emph{et~al.},
  \bibinfo{journal}{New Journal of Physics} \textbf{\bibinfo{volume}{13}},
  \bibinfo{pages}{095015} (\bibinfo{year}{2011}).

\bibitem{sfreservoirs}

\bibinfo{note}{Since the quasi-two-dimensional channel is much longer than
  $\xi$, having superfluid reservoirs changes their compressibility, but does
  not induce macroscopic coherence effects in the channel.}

\bibitem{tempeffects}

\bibinfo{note}{The single particle transport is determined by the energy of
  particles, which in the case of the weakly interacting Fermi gas is set by
  the Fermi energy. Therefore, as long as the gas is degenerate, the conduction
  properties should be close to that of the zero temperature cloud. For the BEC
  case, even at moderate temperatures, the mean energy of particles is set by
  the chemical potential. Except close to the transition where temperature
  induced depletion may be comparable to the disorder induced depletion,
  temperature should have little influence.}

\bibitem{PhysRevB.20.3653}

\bibinfo{author}{\bibfnamefont{L.~N.} \bibnamefont{Smith}} \bibnamefont{and}
  \bibinfo{author}{\bibfnamefont{C.~J.} \bibnamefont{Lobb}},
  \bibinfo{journal}{Phys. Rev. B} \textbf{\bibinfo{volume}{20}},
  \bibinfo{pages}{3653} (\bibinfo{year}{1979}).

\bibitem{PhysRevB.26.1352}

\bibinfo{author}{\bibfnamefont{A.}~\bibnamefont{Weinrib}},
  \bibinfo{journal}{Phys. Rev. B} \textbf{\bibinfo{volume}{26}},
  \bibinfo{pages}{1352} (\bibinfo{year}{1982}).

\bibitem{2007PhRvL..98q0403L}

\bibinfo{author}{\bibfnamefont{P.}~\bibnamefont{{Lugan}}}, \emph{et~al.},
  \bibinfo{journal}{Physical Review Letters} \textbf{\bibinfo{volume}{98}},
  \bibinfo{pages}{170403} (\bibinfo{year}{2007}).

\bibitem{1367-2630-12-7-073003}

\bibinfo{author}{\bibfnamefont{S.}~\bibnamefont{Pilati}}, \emph{et~al.},
  \bibinfo{journal}{New Journal of Physics} \textbf{\bibinfo{volume}{12}},
  \bibinfo{pages}{073003} (\bibinfo{year}{2010}).

\bibitem{PhysRevB.80.104515}

\bibinfo{author}{\bibfnamefont{G.~M.} \bibnamefont{Falco}},
  \bibinfo{author}{\bibfnamefont{T.}~\bibnamefont{Nattermann}},
  \bibnamefont{and} \bibinfo{author}{\bibfnamefont{V.~L.}
  \bibnamefont{Pokrovsky}}, \bibinfo{journal}{Phys. Rev. B}
  \textbf{\bibinfo{volume}{80}}, \bibinfo{pages}{104515}
  (\bibinfo{year}{2009}).

\bibitem{Bourdel:2012fk}
\biburlhref{http://arxiv.org/abs/1210.1096}{%
\bibinfo{author}{\bibfnamefont{T.}~\bibnamefont{Bourdel}},
  \bibinfo{journal}{arXiv e-print 1210.1096}  (\bibinfo{year}{2012}).}

\bibitem{timescale}

\bibinfo{note}{The classical timescale for transport in the speckle pattern
  $\sqrt{m \sigma^2/\bar{V}}$ is below $10$\,$\mu$s for all the data that we
  show, so we can expect to reach a steady state \cite{2011NJPh...13i5015P}}.

\bibitem{2009Natur.460..995G}

\bibinfo{author}{\bibfnamefont{N.}~\bibnamefont{{Gemelke}}}, \emph{et~al.},
  \bibinfo{journal}{Nature} \textbf{\bibinfo{volume}{460}},
  \bibinfo{pages}{995} (\bibinfo{year}{2009}).

\bibitem{PhysRevLett.99.250402}

\bibinfo{author}{\bibfnamefont{G.}~\bibnamefont{Orso}}, \bibinfo{journal}{Phys.
  Rev. Lett.} \textbf{\bibinfo{volume}{99}}, \bibinfo{pages}{250402}
  (\bibinfo{year}{2007}).

\bibitem{1367-2630-13-5-055012}

\bibinfo{author}{\bibfnamefont{L.}~\bibnamefont{Han}} \bibnamefont{and}
  \bibinfo{author}{\bibfnamefont{C.~A. R.~S.} \bibnamefont{de~Melo}},
  \bibinfo{journal}{New Journal of Physics} \textbf{\bibinfo{volume}{13}},
  \bibinfo{pages}{055012} (\bibinfo{year}{2011}).

\end{thebibliography}

\end{document}